\documentstyle[preprint,aps]{revtex}
\tighten
\draft
\begin{document}
\title{Temperature dependent BCS equations \\ with continuum coupling}

\author{
N. Sandulescu$^1$\thanks{On leave from Institute of Atomic Physics, 
Bucharest},
O. Civitarese$^2$
and
R. J. Liotta$^1$,
\\
{\small\it$^{1}$Royal Institute of Technology, 
Frescativ 24, S-10405, Stockholm, Sweden,}\\
{\small\it$^{2}$ Dept. of Physics. Univ. of La Plata,
c.c. 67 1900, La Plata, Argentina.}
}
\date{\today}
\author{%
\parbox{14cm}{\parindent4mm\baselineskip11pt%
{\small
Abstract:
The temperature dependent BCS equations are modified in order to include 
the contribution of the continuum single particle states. 
The influence of the continuum upon the critical temperature 
corresponding to 
the phase transition from a superfluid to a normal state and upon the 
behaviour of the excitation energy and of the entropy is discussed.
}
\vspace{4mm}
}
}
\author{
\parbox{14cm}
}
\maketitle
  
\section{Introduction}

 The effect of temperature upon pairing correlations in nuclei
was studied long ago \cite{Sano}, after the theory of superfluidity
was introduced to describe the appearance of a gap in the low-energy
spectrum of finite nuclei \cite{Bohr}. The extension of the theory 
of superfluidity to the finite temperature case was
prompted  by the search of evidence about the nuclear phase
transition from a superfluid to a normal state, in analogy with
the one found in condensed matter systems \cite{Emery}. 
The vanishing of pairing correlations with temperature is related to
the fact that the excitation energy breaks pairs of particles which block the 
single particle levels close to the Fermi surface, 
where the pairing correlations are initiated.
 
The change of the  pairing properties with the temperature was also studied 
in competition with the angular momentum\cite{Kammuri,Moretto,Goodman}. 
In this case a new effect was observed, which is due to the fact that 
at zero temperature the single-particle states close to the Fermi 
level are already partially blocked by the un-paired nucleons which
form the finite angular momentum of the nucleus. 
Thus, when the temperature is 
switched on the first effect is a depletion  of the partially blocked 
single-particle states which, in turn, induces an enhancement of the pairing 
correlations. 
This behaviour is different for the case of thermal excitations of 
nuclei with zero angular momentum, where the pairing correlations 
decrease monotonically with the increase of the temperature. 
 
The temperature effects on pairing correlations were studied both in the 
BCS \cite{Sano,Kammuri,Moretto,Civ} and HFB \cite{Goodman} approximations, 
usually with a constant single-particle level density approximation. 
The available theoretical evidence shows that for medium and heavy-mass nuclei
the pairing correlations disappear for a critical temperature of the order
of 0.5-1.0 MeV. The excitation energies corresponding to this critical
temperature are quite small for nuclei close to the beta stability line 
and therefore in all such calculations the coupling with the continuum 
spectrum was neglected. 
The situation is different  for nuclei which are far from 
the stability line. In this case the Fermi level lies close to the 
continuum threshold and the coupling with the continuum 
becomes important. We shall focus here on the study of this coupling since,
to our knowledge, the effect of the continuum coupling on thermodynamical 
properties of superfluid nuclei has not been  investigated so far.
 
 The main problem in dealing with the continuum coupling in 
statistical calculations of excited nuclei is the fact that 
the particles moving in the continuum have a finite probability 
to be emitted  from the nucleus. In other words, such processes are
time dependent and they are difficult to accomodate 
in stationary models as BCS or HFB. 
But to extent stationary 
many body theories to  time-dependent
formalisms is not an easy undertaking. In fact it may even be a not 
well  defined  task  since  the  initial  conditions  may  induce 
chaotic solutions.
These  features were already recognized in the beginning of  quantum 
mechanics.  Thus,  Gamow  and Wigner tried  to  reconciliate  the 
outgoing character of the decaying process with the  conveniences 
of  stationarity  by  solving  the  Schr\"odinger  equation  with 
outgoing boundary conditions (for references see \cite{Wigner}). 
The corresponding  solutions are related to the complex poles of the
S-matrix, which define the so-called Gamow resonances.
If the resonances are narrow the real parts of the complex poles of the
S-matrix   give  the  positions   of  the  resonances  while   the
corresponding  imaginary parts give the decay  widths. 
The narrow resonances are very important  to describe  
nuclear correlations, especially in unbound or excited nuclei,  
because  in these states the nucleons could move within the nuclear 
volume during a certain minimum  time, so that they can interact with 
each other. One thus expects that a basis formed by bound and 
narrow Gamow resonances would provide a convenient framework 
to describe decaying 
processes \cite{Liotta,Berggren}. However, the drawback of the calculations
based on this representation is that one gets complex probabilities
which are not always easy to interpret \cite{Berggren}.

Another alternative to describe unstable nuclei is to use a basis 
consisting  of scattering states, instead of Gamow states, 
in the vecinity of the resonant poles \cite{Cbcs,Hfbcs}. 
In  this  case  all quantities are real and  one  does  not  have 
problems of interpreting complex probabilities. Within this
representation  the widths of the resonances are obtained by
evaluating the derivative of the corresponding phase shifts.
This also defines the continuum level density \cite{Beth} 
commonly used to estimate the contribution of the 
continuum to nuclear partition functions 
\cite{Fowler,Tubbs,Kerman,Mosel,Bonche,Dean}. 
The escape of the particles which move in 
resonant states is thus treated in these calculations
as a stationary process, 
reflected by a 
constant particle density at large distances from the nucleus . 
The underling picture \cite{Bonche} is an excited  nucleus in dynamical 
equilibrium with an external nucleonic gas, whose contribution should 
be eventually extracted. 
 Recently a similar framework was used to include continuum 
coupling in BCS equations at zero temperature \cite{Cbcs,Hfbcs}. 
In this paper we will extend this method to study temperature 
dependent BCS equations (TBCS).

\section{Formalism}

The standard procedure to derive the temperature dependent BCS 
equations is to minimize the  grand potential 
corresponding to a pairing Hamiltonian. For a bound 
single particle spectrum with energies $\epsilon_j$ and a 
constant pairing interaction of strength G, the gap and 
particle number equations at a
finite temperature $T$ are given by \cite{Sano},

\begin{equation}\label{eq:gap}
\frac{2}{G}= \sum_{j} \frac{1-2f_j}{2E_j}~,
\end{equation}
\begin{equation}\label{eq:n}
N = \sum_j \frac{1}{2}[1-\frac{\epsilon_j-\lambda}{E_j}(1-2f_j)]
\end{equation}
where $E_j=[(\epsilon_j-\lambda)^2+\Delta^2]^{1/2}$ is the
quasiparticle energy, $f_j=(1+ exp(\beta E_j))^{-1}$ is the Fermi
distribution function and $\beta=1/kT$.
 
In principle the single particle energies $\epsilon_j$  depend
also upon temperature because the average mean field is a function 
of the nuclear excitations. However, in self-consistent temperature dependent 
Hartree-Fock calculations one finds \cite{Bonche} that for 
temperatures below $T=1$ MeV, which is the range explored in
temperature dependent BCS calculations, the single-particle spectrum 
is virtually the same as the one at zero temperature. We will 
also assume here that the single particle energies are 
those at zero temperature.

The contribution of the continuum on thermodynamical properties of
finite nuclei was studied mainly in connection with  the problem 
of  the liquid-gas  phase transition \cite{Mosel,Bonche,Dean}  as 
well as in temperature dependent shell corrections \cite{Tshell}, 
but without including pairing correlations.
 In these calculations the effect of the continuum was introduced into 
the thermodynamical quantities through the level density. 
Thus the grand potential for a non-interacting system was taken as \cite{Dean} 
\begin{equation}\label{eq:gp}
\Omega= - T \int (g_b(\epsilon)+\tilde{g}(\epsilon)) 
          ln(1+exp[-\beta(\epsilon-\lambda)])d\epsilon
\end{equation}
where $g_b(\epsilon)=\sum_j \delta(\epsilon-\epsilon_j)$ is the
level density of the bound spectrum and $\tilde{g}(\epsilon)$ is the 
level density associated with the positive energy spectrum. 
In Ref. \cite{Bonche} it is shown that the grand potential (\ref{eq:gp}) 
describes a nucleus in dynamical equilibrium with a nucleonic gas. 
As discussed above, this is due to the fact that in a stationary 
treatment the nucleons scattered in the continuum are permanently 
emitted from the nucleus. 
To obtain the proper grand potential, i. e. the one  corresponding to the 
nucleus itself, one should take away from Eq. (\ref{eq:gp})  
the contribution of the nucleonic gas. This can be done 
by subtracting from the grand potential (\ref{eq:gp}) the grand potential 
of the free nucleonic gas \cite{Bonche}, or by  
replacing in Eq. (\ref{eq:gp}) the level density $\tilde{g}(\epsilon)$ 
by the quantity  \cite{Dean}
\begin{equation}\label{eq:gc} 
g(\epsilon) =\tilde{g}-g_{free}=
{1\over \pi} \sum_{j}\frac{d\delta_{j}}{d\epsilon} 
\end{equation} 
where $g_{free}$ is the level density in the absence of the mean field 
and $\delta_{j}$ is the phase shift. The quantity $g(\epsilon)$ is  
the continuum level density \cite{Beth}. It takes into account
the contribution of the resonant part of the continuum spectrum.
The continuum coupling can thus be included by replacing in
the grand potential (\ref{eq:gp}) the density $\tilde{g}$ by the 
continuum level density $g(\epsilon)$. 
This is a general recipe which 
can be applied to all quantities derived from the grand potential, as
e.  g. the energy and entropy of the system.

The incorporation  of the continuum in interacting systems
can readly be performed following a similar prescription. 
Thus the contribution of the continuum to the grand potential of an
excited  superfluid nucleus  can be expressed in term of the continuum level 
density  as in Eq. (\ref{eq:gp}), with the difference that now instead of 
the single particle energies one should use the quasiparticle energies. 
The corresponding TBCS equations can be obtained directly 
from Eqs. (\ref{eq:gap}) and (\ref{eq:n}) by replacing the level density 
of the bound states with the total level density, i. e. by
$g_b(\epsilon)+g(\epsilon)$. Thus the TBCS
equations with continuum coupling become,
\begin{equation}\label{eq:gapc}
\frac{2}{G}= \sum_{j} \frac{1-2f_j}{2E_j}+
\int g(\epsilon)
\frac{1-2f(\epsilon)}{2E(\epsilon)} d\epsilon~,
\end{equation}
\begin{equation}\label{eq:nc}
N  =  \sum_j \frac{1}{2}[1-\frac{\epsilon_j-\lambda}{E_j}(1-2f_j)]
+\int g(\epsilon) \frac{1}{2}
[1-\frac{\epsilon-\lambda}{E(\epsilon)}(1-2f(\epsilon))]d\epsilon
\end{equation}
where the second term gives the contribution of the continuum
to the pairing correlations. In the limit T=0 one gets the same 
equations as in Refs \cite{Cbcs,Hfbcs}. For temperatures higher
than the critical temperature the gap vanishes and the particle
number equation is similar to the one used in thermodynamical 
calculations of non-interacting systems 
\cite{Mosel,Bonche,Dean,Tshell}.
  
The contribution of the continuum to the energy and to the entropy can
be introduced in a similar fashion, as mentioned above. One gets, 
\begin{equation}\label{eq:energy}
E = \sum_j n_j \epsilon_j
+\int g(\epsilon) n(\epsilon) \epsilon d\epsilon
-\frac{\Delta^2}{G}~,
\end{equation}
\begin{equation}\label{eq:entropy}
S  =  \sum_j 
\{ ln(\tilde{f}_j)+\beta f_j E_j \} 
+\int g(\epsilon)
\{ ln(\tilde{f}_\nu(\epsilon))+
   \beta f_\nu(\epsilon) E_\nu(\epsilon)\} d\epsilon
\end{equation}
where $n_j$ is the occupancy of the state of energy $\epsilon_j$, given by
\begin{equation}\label{eq:occ}
n_j=
\frac{1}{2}[1-\frac{\epsilon_j-\lambda}{E_j}(1-2f_j)]
\end{equation}
and $\tilde{f}_j=1+exp(-\beta E_j)$. A similar notation is used for 
the corresponding quantities in the continuum.

In the limit of vanishing widths the continuum level density becomes 
a sum of Dirac delta functions and the resonances act as bound
states ("quasibound states"). This is the case for, e. g.,  protons 
trapped in a high Coulomb barrier. 
If a resonance is not narrow then a pair scattered 
to that resonance have a large probability to escape from the 
system. Therefore, its contribution to the pairing 
correlation is small as compared with the corresponding 
contribution from a quasibound state with similar energy and angular 
momentum \cite{Cbcs,Hfbcs}.
  
Although narrow resonances play a fundamental role in the 
enhancement of pair correlations (and in most other measurable
physical processes as well) their proper inclusion in the 
applications may be difficult. The reason for this is that
in the region of a narrow resonance the level density increases
abruptly and the numerical evaluation of the integrals in the equations 
above may require extremely small mesh intervals. One can 
circumvent  this problem by noticing that the resonance is 
narrow  because there is a pole of the $S$ matrix which  is  very 
close to the real energy axis. Therefore, one can 
evaluate the integrals by changing the integration path, that 
is by choosing a contour $C$ in the complex energy plane which 
embodies the real axis around the
narrow resonance,  and by
thereafter applying the Cauchy theorem 
\cite{Cbcs,Tshell}. Eq. (\ref{eq:energy}) thus becomes
\begin{equation}\label{eq:enc}
E = \sum_{i}n_i\epsilon_i +
\sum_{\nu}n({\bf E_\nu}){\bf E_\nu}
+\int_{C} g(\epsilon)n(\epsilon)\epsilon d\epsilon
-\frac{\Delta^2}{G}
\end{equation}
where $n({\bf E_\nu})$ are the occupation probabilities calculated in 
the complex poles $ {\bf E_\nu} $ enclosed by the path $C$.
If one neglects the contribution of the integral 
over the contour $C$ in Eq. (\ref{eq:enc}) then the energy $E$
would become complex. This is the case in representations based on Gamow 
resonances  only  \cite{pat}. As already mentioned,  within  such 
representations one obtains complex physical quantities which have 
to be interpreted. For the energy such a task is rather easy, since 
already by looking at the temporal evolution of the wave function 
one realizes that the real part corresponds to the actual energy 
of the system while the imaginary part is related to the 
corresponding decay probability. This interpretation is valid
if the resonance is narrow,  i. e. if the ratio between the width
and the energy  of the resonance is small \cite{kuku,Tore}. But 
the interpretation of complex probabilities in general is not so 
straightforward \cite{Berggren,pat}. These problems do not appear
here because in Eq. (\ref{eq:enc}) the contribution of the integral 
over the contour $C$ is included and therefore the energy $E$ is real.

\section{Numerical application}

 In order to illustrate how the continuum affects the properties of 
superfluid nuclei close to the drip line, in what follows 
we present the results given by the TBCS equations 
for the isotope $^{84}Ni$. 

We calculated the single particle spectrum by using the HF 
approximation and the Skyrme III interaction \cite{SIII}.
The resonant energies are defined as the energies where the 
phase shift passes through $\pi/2$ with a positive slope 
\cite{kuku}. The width is extracted 
from the value of the energy where the 
derivative of the phase shift 
is half of its maximum value. 
The bound states outside the closed shell N=50 and the resonant states 
considered in the TBCS calculations are listed in Table 1. It can be
seen that these states form the equivalent of the major  shell 
N=50-82. One notices that the relative 
positions of the single-particle states differ substantially with
the ones corresponding to beta stable nuclei. Thus, the states with 
low angular momenta are shifted down  as compared with the ones
with high angular momenta. This is
a general feature related to the diffusivity of
the mean field in nuclei close to the drip line \cite{Hamamoto}, 
which is larger than the corresponding one in stable nuclei .

The TBCS equations (\ref{eq:gapc}) and (\ref{eq:nc}) are solved 
starting with the HF spectrum calculated at zero temperature. In the 
absence of experimental information on heavy $Ni$ isotopes 
in the open shell $N=50-82$ (the heaviest known $Ni$ isotope 
is the double magic nucleus $^{78}$Ni) we use for the strength of the
pairing force the standard value $G=25/A$ MeV, where $A$ is the 
mass number. 
 
The variation of the gap with temperature is shown in Fig. 1.
In the limit of $T=0$ the gap has the value $\Delta(0)=0.955$ MeV and
decreases to zero at the critical temperature $T_c=0.524$ MeV. 
This critical temperature is smaller by about $3.8\%$ than
the one obtained from the relation $T_c=0.57\Delta(0)$,
which is the value predicted by a constant level density approximation 
\cite{Sano}.

In Fig. 1 it is also shown the gap one would get if 
the width of the resonances are neglected, i.e. if in 
Eqs. (\ref{eq:gapc}-\ref{eq:nc}) one would take instead of the continuum 
level density a sum of Dirac delta functions. The gap at zero temperature 
is in this case bigger than in the previous case \cite{Cbcs,Hfbcs}. 
As discussed above, the wider is a resonance the smaller is its
contribution to 
the pairing correlation. That is, a pair scattered to a wide resonance 
spends less time in that state as compared to the time that the 
same pair would spend in a quasibound state and, therefore,
the wider resonance contributes less to the pairing correlations. 
This has the important consequence that neglecting the widths  of 
the resonances one increases the correlations of the system and 
therefore increases the critical temperature. This is what 
happens, e. g., with  calculations that  quantize  the 
continuum by using an impenetrable box if the dimensions of the
box are not extremely large (so that one gets a dense enough 
spectrum in the energy regions of resonant states).  
However, one has to mention that this is not a 
serious problem if the resonances are very narrow.

Neglecting the
widths one gets for the critical temperature the value $T_c=0.722$ 
MeV, which is only $1\%$ larger than the value one would obtain in 
a constant level density approximation. This indicates that 
this approximation would work quite well for nuclei close to the
proton drip line, where the width of the proton single-particle 
resonances may be so narrow that they can be neglected. 
 
In Fig. 2 we show the dependence of the excitation energy upon 
temperature up to the point where the superfluid phase vanishes.
It can be seen that if the  effect of
the width of resonant states is neglected, then the slope of the 
excitation energy becomes much  smaller. This is also a manifestation 
of the fact that by neglecting the widths, the ground state 
becomes more correlated and therefore stiffer to thermal excitations.
The same effect is observed for the entropy, as seen in Figure 3.  

At critical temperature the excitation energy has the value 2.185 MeV 
and the entropy is 5.708. For a constant level 
density $g$ the excitation energy at critical temperature is given
by \cite{Sano} $ E_c\approx aT^2_c+0.5g\Delta(0)^2$, 
where $a=\frac{\pi^2}{3}g$ is the level density parameter, which in 
this case acquires the value $a=3.958$ MeV$^{-1}$.  
We found that neglecting the widths the level density parameter
increases to the value $a=4.143$ MeV$^{-1}$. 
This implies that the difference in 
the excitation energies seen in Fig. 2 is esentially due to the mean 
occupancy of the resonant states (which is larger if the widths are 
neglected) and not due to the effective level density parameter.

In conclusion, in this paper we have extended the temperature dependent BCS 
equations by introducing the coupling with the single-particle 
continuum. The contribution of the continuum is given  by  the resonant 
states and their effect is taken into account through the continuum 
level density. 

We  found that the widths of the resonances affect  significantly 
all physical quantities. In particular, the pairing correlations are 
diminished and this modifies significantly the value of the
critical temperature at which the supefluid phase disappears.
Also the dependence upon temperature of the excitation energy and 
of  the entropy is considerably affected by the  widths of 
the resonances, i.e. by their lifetime. 

However, in the case of proton superfluidity the 
single-particle resonant states close to the continuum 
threshold may have a very small width and 
therefore they can be treated in the TBCS calculations as quasibound
states, neglecting their widths altogether.

(O.C.) is partially supported by the grant (PICT0079) and by the 
CONICET, Argentina.
(N.S.) is supported by the Wenner-Gren Foundation.

\newpage

\mediumtext
\begin{table}
\caption{
Neutron single particle states corresponding to $^{84}$Ni.
These are the states used  in  the  TBCS calculation.  $E_n$ 
is the energy and 
$\Gamma_n$ the width of the state labelled by $n$. 
Both quantities are in MeV.
\label{table1}}
\begin{tabular}{lrr}
n & $E_n$ & $\Gamma_n$\\
\tableline
$2d_{5/2}$   & -1.470  & ----- \\
$3s_{1/2}$   & -0.730  & ----- \\
$d_{3/2}$   &  0.486  & 0.112 \\
$g_{7/2}$   &  1.605  & 0.010 \\
$h_{11/2}$  &  3.296  & 0.016 \\
\end{tabular}
\end{table}
         
\section*{Figure Captions}

Figure 1: Dependence of the gap parameter $\Delta$ upon the temperature $T$.
          The dashed line corresponds to the case when the resonant states
          are considered as quasibound states.
 
Figure 2: Excitation energy plotted as a function of temperature. 
          The dashed line corresponds to the case when the resonant states
          are considered as quasibound states. The energy is plotted up to
          the critical temperature.

Figure 3: Entropy plotted as a function of temperature.  
          The dashed line corresponds to the case when the resonant states
          are considered as quasibound states. The entropy is plotted up to
          the critical temperature.


\newpage

\begin{references}
\bibitem{Sano}
{M. Sano and S. Yamasaki, Prog. Theor. Phys., {\bf 29} (1963) 397.}
\bibitem{Bohr}
{A. Bohr, B. Mottelson and D. Pines, Phys. Rev. {\bf 110} (1958) 936.}
\bibitem{Emery}
{V. J. Emery and A. M. Sessler, Phys. Rev. {\bf 119} (1960) 43.}
\bibitem{Kammuri}
{T. Kammuri, Prog. Theor. Phys., {\bf 31} (1964) 595.}
\bibitem{Moretto}
{ L. G. Moretto, Nucl. Phys.{\bf A 185} (1972) 145.}
\bibitem{Goodman}
{A. L. Goodman, Nucl. Phys.{\bf A 352} (1981) 30.}
\bibitem{Civ}
{O. Civitarese, G. G. Dussel and R. P. J. Perazzo, Nucl. Phys.
{\bf A 404} (1983) 15}
\bibitem{Wigner} 
{T. Teichman and E. P. Wigner, Phys. Rev. {\bf 87} (1952) 123.}
\bibitem{TP}
{T. Berggren and P. Lind, Phys. Rev. {\bf C47} (1993) 768.}
\bibitem{Liotta}
{R. J. Liotta, E, Maglione, N. Sandulescu and T. Vertse,
Phys. Lett. {\bf B367} (1996) 1.}
\bibitem{Berggren}
{T. Berggren, Phys. Lett. {\bf B73} (1978) 389.}
\bibitem{Cbcs}
{N. Sandulescu, R. J. Liotta, and R. Wyss,
Phys. Lett. {\bf B394} (1997) 6.}
\bibitem{Hfbcs}
{N. Sandulescu, Nguyen Van Giai, and R. J. Liotta, nucl-th/9811037}
\bibitem{Beth}
{E. Beth and G. E. Uhlenbeck, Physica {\bf 4} (1937) 915.}
\bibitem{Fowler}
{W. A. Fowler, J. M. Lattimer and G. E. Brown, Astrophys. J.
{\bf 226} (1978) 984.}
\bibitem{Tubbs}
{D. J. Tubbs and S. E. Koonin, Astrophy. J. {\bf 232} (1979) L59. }
\bibitem{Kerman}
{A. K. Kerman and S. Levit, Phys. Rev. {\bf C24} (1981) 1020.}
\bibitem{Mosel}
{U. Mosel, Phys. Lett. {\bf B 46} (1973) 8 }
\bibitem{Bonche}
{P. Bonche, S. Levit and D. Vautherin, Nucl. Phys.{\bf A 427} (1984) 278.}
\bibitem{Dean}
{D. R. Dean and U. Mosel, Z. Phys. {\bf A 322} (1985) 647}
\bibitem{Tshell}
{N. Sandulescu, O. Civitarese, R. J. Liotta, and T. Vertse,
Phys. Rev. {\bf C55} (1997) 1250.}
\bibitem{pat} 
{P. Curutchet, R. J. Liotta and T. Vertse, 
Phys. Rev. {\bf C 39} (1989) 1020.}
\bibitem{kuku} 
{V. I. Kukulin, V. M. Krasnopolsky and J. Horacek, "Theory of 
Resonances", Kluwer Academic Publishers, London, 1989,
Chapter 2.}
\bibitem{Tore} 
{T. Berggren, Phys. Lett. {\bf B 373} (1996) 1.}
\bibitem{SIII}
{M. Beiner, H. Flocard, Nguyen Van Giai, and P. Quentin, 
Nucl. Phys. {\bf A238} (1975) 29. }
\bibitem{Hamamoto}
{I. Hamamoto, H. Sagawa and X. Z. Zhang, 
Phys. Rev. {\bf C53} (1996) 765.}
\end{references}
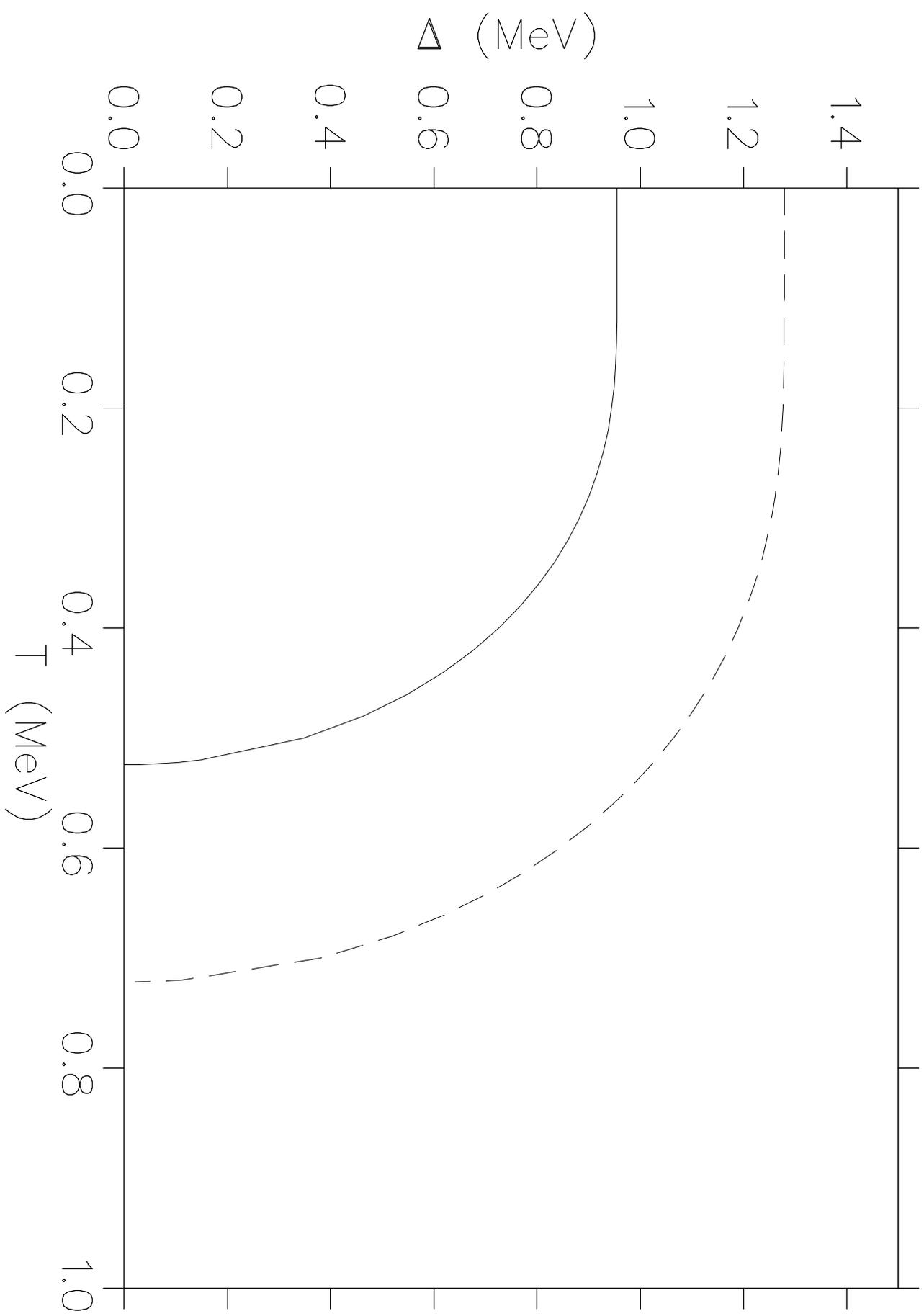
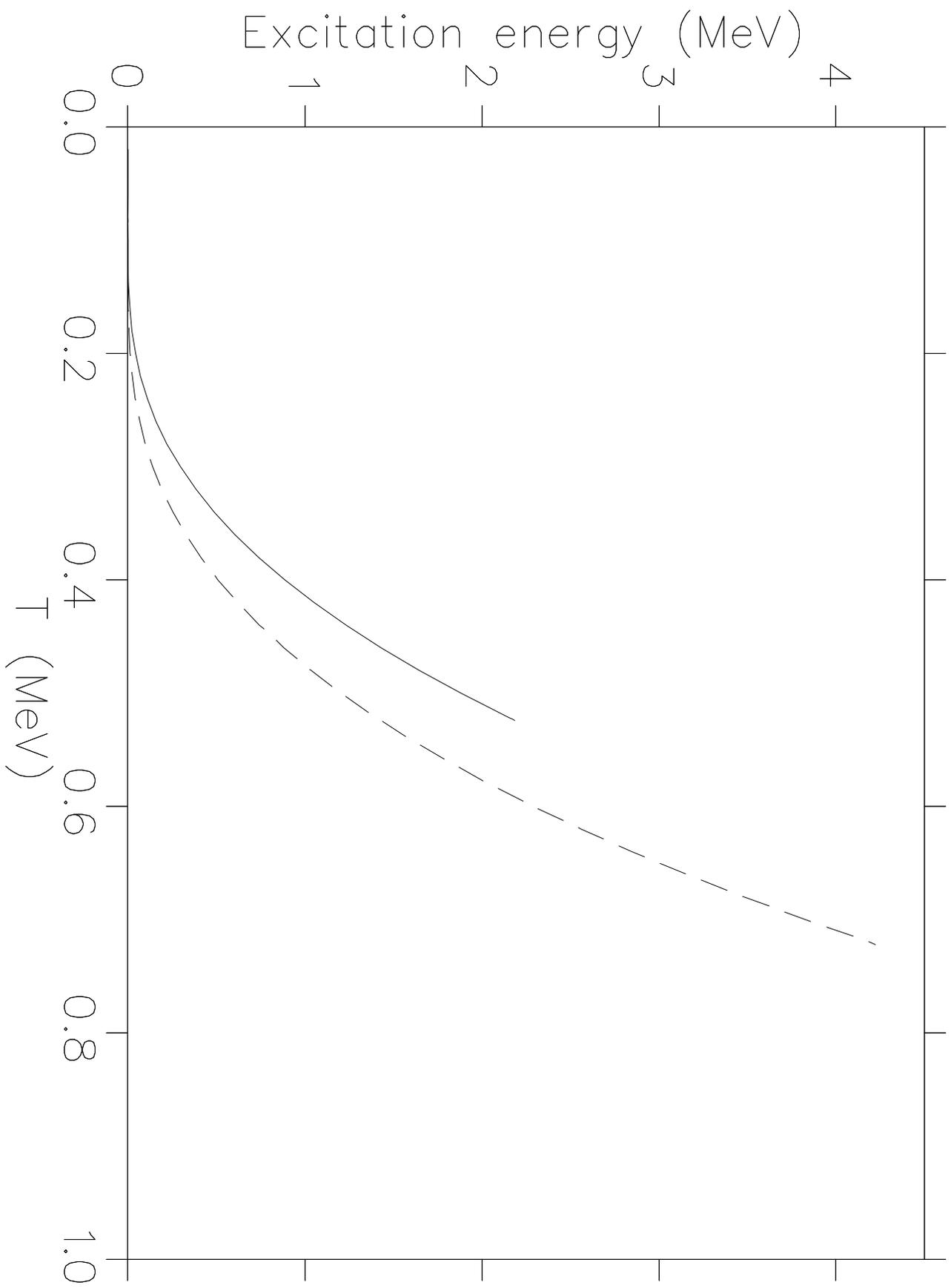
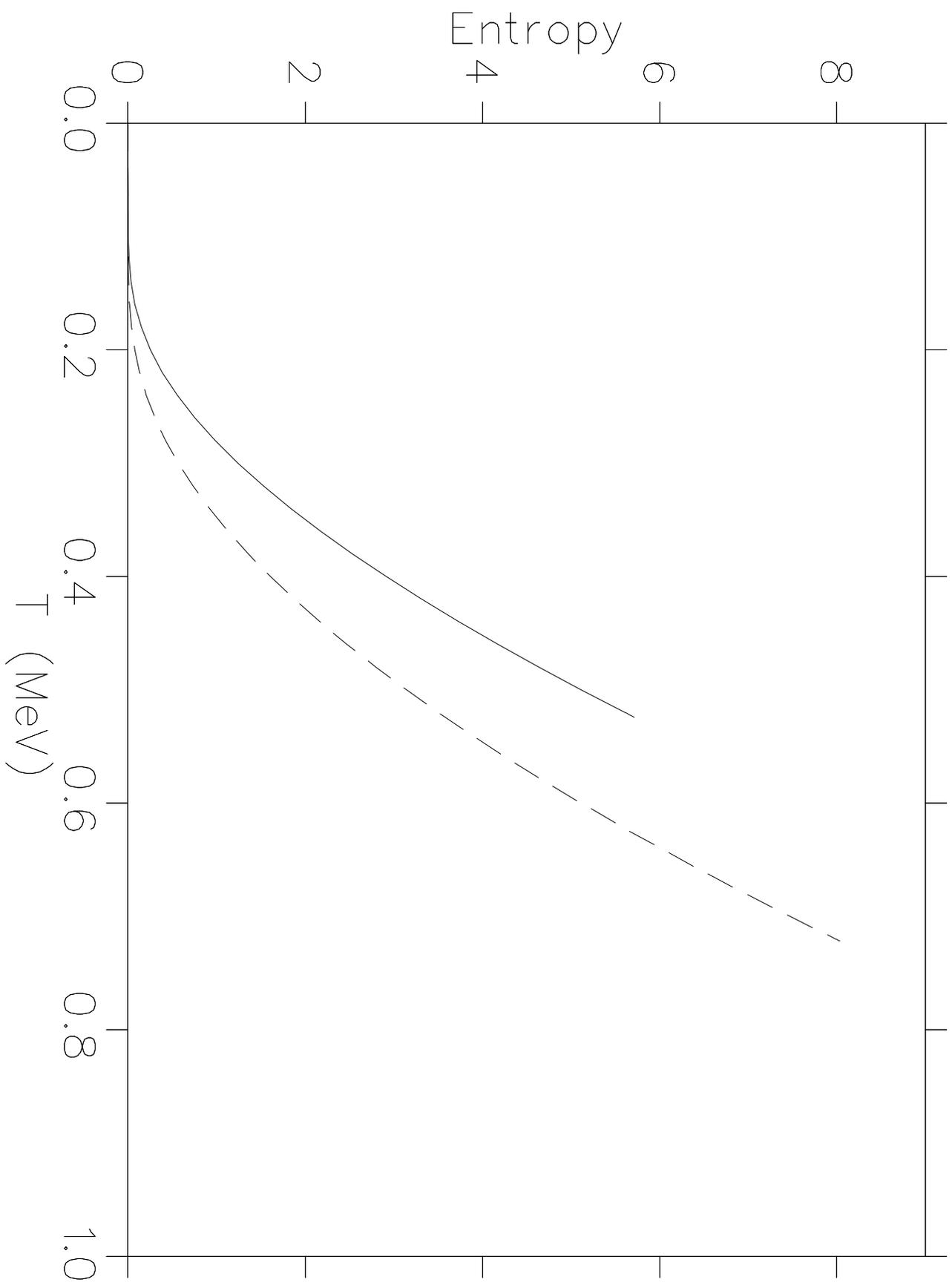
\end{document}